\documentclass{mem}
\usepackage{natbib}\usepackage{txfonts}\usepackage{balance}
\usepackage{graphicx}
\usepackage[a4paper]{hyperref}
\idline{76}{1}

\newcommand{\Acta}[3]{{#1}, Acta Astron., \vol{{#2}}, {#3}.}
\newcommand{\AJ}[3]{{#1}, AJ, \vol{{#2}}, {#3}.}
\newcommand{\ApJ}[3]{{#1}, ApJ, \vol{{#2}}, {#3}.}
\newcommand{\ApJS}[3]{{#1}, ApJS, \vol{{#2}}, {#3}.}

\newcommand{\AandA}[3]{{#1}, A\&A, \vol{{#2}}, {#3}.}

\newcommand{\MNRAS}[3]{{#1}, MNRAS\rm, \vol{{#2}}, {#3}.}

\newcommand{\vol}[1]{{\mbox{#1}}}

\newcommand{\Mv}{\mbox{$M_{V}\,$}}
\newcommand{\Mk}{\mbox{$M_{K}\,$}}
\newcommand{\FeH}{\mbox{[Fe/H]}\,}

\newcommand{\mM}{\mbox{$(m-M)$}\ }

\begin{document}
\def\teff{$T\rm_{eff }$}
\def\kms{$\mathrm {km s}^{-1}$}

\title{
How good are RR Lyrae and Cepheids really as Distance Indicators?
}

   \subtitle{The observational approach}

\author{
Jesper Storm
          }


\institute{
Astrophysikalisches Institut Potsdam --
An der Sternwarte 16,
14482 Potsdam,
Germany
\email{jstorm@aip.de}
}

\authorrunning{Storm}

\titlerunning{RR Lyrae and Cepheids as Distance Indicators}

\abstract{
A number of recent technical developments, including the Hipparcos
satellite, the Hubble Space Telescope fine guidance sensors and long
base line near-IR interferometry has made it possible to employ
several largely geometrical methods to determine direct distances to
RR Lyrae stars and Cepheids. The distance scale now rests on a much firmer
basis and the significant differences between the distances based
on RR Lyrae stars (short) and Cepheids (long) to the LMC have been largely
eliminated. The effects of metalicity on the RR Lyrae period-luminosity
(PL) relation in the $K$-band as well as on the Cepheid PL relation appears
to be the main remaining issues but even here empirical results are
beginning to show convergence. I review here some of these recent developments
seen from the perspective of the near-IR surface brightness method.

}
\maketitle{}
%

\section{Introduction}

  RR Lyrae stars and Cepheids are intermediate mass pulsating stars which are
of fundamental importance for the calibration of the extragalactic
distance scale. Recently significant progress has been made in obtaining
largely geometric calibrations of these standard candles and a well
founded consensus
is finally starting to emerge regarding the distance to the Large
Magellanic Cloud which provides the first step on the distance ladder as 
employed e.g. by the Hubble Space Telescope Key Project on the 
Extragalactic Distance Scale (Freedman et al. \citealp{Freedman01}).

%

\section{The RR Lyrae stars}
The RR Lyrae stars are valuable standard candles via the $\FeH-\Mv$
and $\log P - \Mk - \FeH$ (PLZ(K)) relations. The first relation is the
most widely used as it is based on optical observational data.
Unfortunately the relation exhibits a significant scatter and can only
be considered an ensemble relation. Also the slope of the relation has
been a contentious issue ranging from 0.16 to 0.30 (e.g. Jones et al.
\citealp{Jones92}, Sandage \citealp{Sandage93}) and the
relation might not even be linear (see e.g. Caputo et al. \citealp{Caputo00}).
The second relation, first discovered observationally by Longmore
et al. \cite{Longmore86} and \cite{Longmore90}, is potentially
much more powerful as it exhibits a very low intrinsic scatter. D'Allora
et al. \cite{Dallora04} observed an intrinsic scatter of only 0.026~mag
for the (equal abundance) RR Lyrae stars in the LMC star cluster
Reticulum. Recent theoretical work however, suggests that the metalicity effect
is quite pronounced, of the order 0.2~mag/dex.  Bono et al. \cite{Bono03}
finds:
\begin{equation}
\label{eq.PLZk}
\Mk = -0.77-2.101\log P+0.231\FeH
\end{equation}
while Catelan et al. \cite{Catelan04} finds:
\begin{equation}
\Mk = -0.597-2.353\log P+0.175\log Z
\end{equation}

We have work in progress to observationally constrain this relation.

\subsection{Trigonometric Parallax}

The purest geometrical method which we have at our disposal for measuring
distances to stars is to measure the trigonometric parallax.  Hipparcos
and HST have expanded very significantly the volume in space where this
is feasible.  Even so Hipparcos could barely reach one RR Lyrae star,
namely RR Lyr itself, resulting in a fairly uncertain parallax measurement
of $4.38\pm0.59$~mas, Perryman et al. \cite{Perryman97}. More recently
Benedict et al. \cite{Benedict02a} using the fine guidance sensor on HST
found a value of $3.82\pm0.20$~mas, corresponding to $\Mv=0.61\pm0.1$
and $\Mk = -0.56\pm0.1$ for an assumed absorption of $A_V=0.07$. This
is in excellent agreement with the value predicted by Eq.\ref{eq.PLZk}
of $-0.57$. Due to the intrinsic width of the $\FeH-\Mv$ relation
this measurement can only constrain the zero point to within 0.13~mag
(Cacciari and Clementini, \citealp{CC03}).

\subsection{Statistical Parallax}
The statistical parallax method assumes a dynamical model for the sample
of stars being analyzed assuming that they belong to a dynamically well
defined sample. Modern results from Layden et al. \cite{Layden96}
($\Mv=0.71$ for $\FeH=-1.61$), Gould and Popowski \cite{Gould98} ($\Mv =
0.77$ for $\FeH = -1.6$)
tend to support a faint magnitude of RR Lyrae stars when compared to
other methods mentioned here. Applied to the LMC these results still
support a short distance modulus. However, the method seems more and
more isolated which suggests that the method might still suffer from
systematic errors possibly related to the adopted models or due to 
unresolved biases in the observed samples.

%
\subsection{Baade-Wesselink type analysis}

  The Baade-Wesselink type analysis gives accurate individual distances
and absolute magnitudes to pulsating stars like RR Lyrae and Cepheids.
Jones et al. \cite{Jones92} derived a $\Mv-\FeH$ for field RR Lyrae
stars with a rather shallow slope. More recently Fernley et al.
\cite{Fernley98} reanalyzed the available data and found
\begin{equation}
\Mv = 0.20(\FeH - 1.5) + 0.68
\end{equation}

  For RR Lyr with a metalicity of $\FeH=-1.39$ this gives $\Mv = 0.70$
and for the LMC adopting $\FeH=-1.5$ and $\langle V_0 \rangle = 19.07$
from Clementini
et al. \cite{Clementini03}  this leads to $(m-M)_0 = 18.39$. However,
Cacciari et al. \cite{Cacciari00} using revised model atmospheres etc
for RR~Cet found that the stars should be brighter by about 0.1~mag
bringing these results very much into line with the canonical LMC
distance (see Sec.\ref{sec.LMC}) as well as the trigonometric parallax
result for RR Lyr itself.

  More recently Kovacs \cite{Kovacs03} applied the Baade-Wesselink
method using new model atmospheres from
Castelli et al. \cite{Castelli97}. He found very good agreement with the
implicit temperature scale from Fouqu\'e and Gieren \cite{FG97} from the
near-IR surface-brightness method. He also found good agreement between
the RR Lyr and Cepheid distance scales to the LMC and found a best
estimate of $(m-M)_0 = 18.55$, which is quite different from the short
distance implied by the analysis by Jones et al. \cite{Jones92}.

\subsection{ZAMS fitting to sub-dwarfs}

The classical ZAMS fitting to globular clusters has recently been
revisited by Gratton et al. \cite{Gratton03}. They have used local
sub-dwarfs with accurate Hipparcos parallaxes to determine distances
to globular clusters by main sequence fitting in a very careful
analysis. They find a relation
\begin{equation}
\Mv = 0.22(\FeH+1.5)+0.56
\end{equation}

  Combining this with the observed $\langle V_0 \rangle=19.07$
for the LMC RR Lyrae
stars from Clementini et al. \cite{Clementini03} leads reassuringly
to the canonical LMC distance of $18.51\pm0.09$.

\section{The distance to the LMC}
\label{sec.LMC}

  The distance to the LMC is used by the HST Key Project as the stepping
stone to the extragalactic distances. They adopted a value of
$18.50\pm0.1$.  Recent reviews based on non-Cepheid distance estimates
all tend to be in good agreement with this value. Walker \cite{Walker99}
found a value of $18.55\pm0.1$. Benedict et al. \cite{Benedict02a}
averaged results from 80 recent studies using 21 different methods and
found $18.47\pm0.04$, albeit with a large spread. 
Tammann, Sandage, and Reindl \cite{Tammann03}
found $18.54\pm0.02$ based on 13 studies. Cacciari and Clementini
\cite{CC03} found $18.48\pm0.05$ from 10 studies based only on RR Lyrae
stars. Consequently it appears that the distance to the LMC is close
to the value of $18.50$ and methods which consistently gives results
different from this value are likely to suffer from yet undetected
systematic errors.

\section{The Cepheids}

\subsection{The near-IR surface brightness method}

  The near-IR surface-brightness (ISB) method is a variant of the Barnes-Evans
(Barnes and Evans, \citealp{BarnesEvans76}) Baade-Wesselink type analysis. It
has recently been calibrated by Fouqu\'e and Gieren \cite{FG97} using
interferometric measurements of non-pulsating giants and super giants.
Gieren et al. \cite{Gieren98}, applying this calibration to a sample
of Galactic Cepheids, found that the zero point agreed well with
that found for galactic open cluster Cepheids from ZAMS fitting.
However, the slope of the derived galactic Cepheid PL relation was
significantly steeper than observed in the LMC. If this effect is real
it will have serious consequences for the application of the Cepheid PL
relation as a distance indicator.

Most recently the method was applied to
a sample of SMC Cepheids (Storm et al. \citealp{Storm04}) and LMC Cepheids
(Gieren et al. \citealp{Gieren05}, Storm et al. \citealp{Storm05}). 
Storm et al. \cite{Storm05}, using a sample of Cepheids in the LMC
star cluster NGC1866, determined the random error per star to be
only 0.11~mag, somewhat larger than the estimated errors from the
method. Barnes et al. \cite{Barnes05} have made a complete Bayesian statistical
analysis propagating the errors through the method and they also find
random errors of this order, suggesting that the method provides quite
accurate distances to individual stars.

Gieren et al. \cite{Gieren05} showed that the ISB method applied to
Galactic and LMC Cepheids, respectively, results in slopes which are
indistinguishable, suggesting that the slope of the PL relation is at
most very weakly dependent on the metalicity of the sample. However,
the slopes remain different from the directly observed slopes in the LMC
from OGLE2 (Udalski et al.  \citealp{Udalski99}) (as modified by Fouqu\'e
et al. \cite{FSG03}) and Persson et al.  \cite{Persson04}. Gieren
et al. \cite{Gieren05} (this volume) showed that the ISB distance
estimates to the LMC Cepheids are period dependent, which of course
is non-physical. The data is not yet entirely conclusive as the short
period stars are all members of the cluster NGC1866. However, taken at
face value they suggest that the conflict can be resolved by adopting
a stronger period dependence of the project factor $p$ which is used to
convert the observed radial velocities into pulsational velocities. These
pulsational velocities are used in the ISB method to determine the stellar
distance and radius. The physical understanding of this effect is still
lacking but one can consider the empirical adjustment of the $p$-factor as
a way to parameterize the problem and thus reconcile all the observational data.


\subsection{Interferometry}
Very recently it has become possible to directly measure angular
diameters of Cepheids. Nordgren et al \cite{Nordgren02} reports
results for 3 Cepheids using a number of interferometers, and Kervella
et al. \cite{Kervella04a} reports results for 7 Cepheids using the
VLTI. For the largest stars they even measure the angular diameter
variation with phase. These results are truly impressive and provides
fundamental geometrical constraints on the distance scale. Kervella
et al.  \cite{Kervella04b} have compared the results from interferometry
with results from the ISB method for the largest (in an angular diameter
sense)
Cepheid in the sky, $\ell$~Car.  The agreement is very good and with the
rapidly expanding body of observational data we can already now employ
the calibration of the surface-brightness relation from interferometry
on Cepheids in the ISB technique.  This means that we no longer have to
rely on the assumption that the pulsating and non-pulsating stars follow
the same relation, even though it turns out to be a fairly good assumption.

  It is here important to remember that to derive distances and radii
for the Cepheids the interferometric method also relies on the conversion
of radial velocity data into pulsational velocities, just like the ISB
method. Thus if the $p$-factor is period dependent as suggested by Gieren
et al. \cite{Gieren05} then this would also have important consequences
for the interferometrically determined distances and radii.

%
%

\subsection{Trigonometric parallax}
  Hipparcos has measured trigonometric parallaxes for a large number of
Cepheids (Feast and Catchpole \cite{FC97}). The individual errors are
rather large but careful analysis results in a zero point which
corresponds to an LMC distance of 18.7. Fouqu\'e et al. \cite{FSG03}
revised the result to 18.50 by adopting the same LMC PL relation and
reddening scale as employed by the HST key Project.

  More recently Benedict et al. \cite{Benedict02b} have measured the
trigonometric parallax to $\delta$~Cep using the HST fine guidance
sensor. They found a value of $\pi_{\mbox{\scriptsize abs}} = 3.66 \pm
0.15$~mas which corresponds to $\Mv=-3.54$ for a visual extinction of
$A_V=0.30$ as adopted by Storm et al.  \cite{Storm04}). This compares
reasonably well with the ISB result from Storm et al. \cite{Storm04}
of $\Mv=-3.43$ but even better with the result after correcting the
$p$-factor of $\Mv=-3.59$ from Gieren et al.  \cite{Gieren05}.

\subsection{ZAMS fitting and the Pleiades}
  Zero Age Main Sequence (ZAMS) fitting to open star clusters containing
Cepheids using the Pleiades cluster or a theoretical ZAMS as a
template was for a long time the most direct way to determine the
zero point of the Cepheid PL relation. The Hipparcos satellite could
tie this calibration in with a geometrical measure of the distance to
the Pleiades. Unfortunately the Hipparcos measurements resulted in a
very short distance, $\mM=5.37\pm0.06$  (van Leeuwen \cite{Leeuwen99})
disagreeing substantially with previous measurements. This result provoked
a surge of investigations to understand whether the Hipparcos result
was wrong or that the understanding of the ZAMS fitting method itself
was seriously flawed. A number of fundamental and largely geometrical
distance determination methods have now been employed to objects in
the Pleiades. Most recently Soderblom et al.  \cite{Soderblom05}
have obtained parallaxes to three Pleiades stars using the HST
fine guidance sensors finding a value of $\mM=5.65\pm0.05$. These
results agree well with recent ZAMS fitting results: $5.60\pm0.04$,
Pinsonneault et al. \cite{Pinsonneault98}, $5.61\pm 0.03$ Stello and
Nissen \cite{Stello01}, $5.63\pm0.05$ Percival et al. \cite{Percival05}
as well as results from an eclipsing binary, $5.60\pm 0.05$, Munari
et al. \cite{Munari04}. Dynamical parallaxes for Atlas by Pan et
al. \cite{Pan04} and Zwahlen et al. \cite{Zwahlen04} further confirm
this. Consequently there now seems to be a consensus that the ZAMS fitting
scale as most recently propagated by Turner and Burker \cite{Turner02}
stand. van Leeuwen and Fantino \cite{Leeuwen05} have now re-reduced the
complete Hipparcos dataset and the new results are eagerly awaited.

%

\subsection{The effect of metalicity on the PL relation}

  It has long been suspected that the Cepheid PL relation might be
affected by metalicity, but conclusive empirical evidence has remained
elusive for a long time. Both on the slope and the
zero-point of the relation can be affected.

  Gieren et al. \cite{Gieren98} found a significantly different slope
between the Galactic and LMC samples of Cepheids based on ISB analysis
of Galactic Cepheids. This result has recently been revised by Gieren et
al. \cite{Gieren05} after applying the ISB method directly to LMC
Cepheids. This result is not conclusive yet, awaiting the analysis
of a larger sample of LMC Cepheids.
Tammann, Sandage and Reindl \cite{Tammann03} based partly on the Gieren
et al. \cite{Gieren98} results also found a significantly different
slope between the LMC and the Galactic Cepheids. On the other hand
Udalski et al. \cite{Udalski01} did not find a significant difference
between the low metalicity sample in IC1613 ($\FeH=-1.0$) and the 
LMC ($\FeH = -0.5$) suggesting that at least in this metalicity range
there is not a significant effect. 

  The possible effect on the zero-point has also been elusive for a
long time, but recent empirical data seems largely to agree on the sign
and approximate size of the effect. As the effect is rather small
it is also very difficult to measure. The best constraints can be
obtained for PL relations with
low intrinsic scatter and with a weak sensitivity to reddening, which
of course are also the relations which are most useful for extragalactic
distance determination. Currently these requirements are best met by the
PL relation in the Wesenheits index as adopted by the
Key Project and the $K$-band PL relation. It should be noted that the
effect is likely to differ in different photometric bands.

The Key Project adopted a value of $-0.20\pm0.2$~mag/dex for the
metalicity sensitivity of the PL relation in the Wesenheits index in the sense
that metal-rich Cepheids are brighter. This choice was largely based
on the measurements presented by Kennicutt et al. \cite{Kennicutt98}.
These results have been confirmed more recently by e.g. 
Groenewegen et al. \cite{Groen04} ($-0.27\pm0.08$~mag/dex) from
five MC Cepheids and 37 galactic Cepheids with individual metalicity
measurements, by Storm et al. \cite{Storm04} ($-0.29\pm0.19$~mag/dex)
from a differential analysis based on the ISB analysis of SMC and
Galactic Cepheids\footnote{The conclusions regarding the metalicity
effect in that paper are based on a purely differential measurement
which is independent of any probable revisions of the $p$-factor, so
this result stands even if the $p$-factor has a stronger dependence on
luminosity.}. Sakai et al. \cite{Sakai04} has performed an analysis of
Cepheids in 17 galaxies with distances from the TRGB method and finds an
effect of $-0.24\pm0.05$.


%
%
%
  Recent empirical investigations seems to agree, with a few exceptions,
that the metalicity effect is fairly weak ($-0.25\pm0.1$~mag/dex)
and in the sense that metal rich Cepheids are brighter. The decisive
measurement is still missing and the effect might well by non-linear and
also depend on other chemical elements, in particular helium, as suggested by
theoretical work of Fiorentino et al. \cite{Fiorentino02}. 	


\section{Conclusions}
The distance to the Large Magellanic Cloud is converging on a value close
to $18.50\pm0.10$ as adopted by the HST Key Project. Methods which give
results significantly different from this value deserve closer scrutiny
as there might be some important astrophysics to be learned.

  The Cepheid PL relation remains the pivotal distance estimator and the
effect of metalicity on the relation is the main remaining
uncertainty. The issue of the supposedly different PL slopes between LMC and
Galactic Cepheids might soon be fully resolved and the effect on the
zero-point is similarly, at least from an empirical point of view,
largely converging on a value not far from that adopted by the HST Key
Project.

%

\bibliographystyle{aa}

\end{document}